\documentclass[twoside,a4paper]{article}
\usepackage{abstract}
\usepackage[utf8]{inputenc}
\usepackage{graphicx}
\graphicspath{{./images/} }
\usepackage[font=small,labelfont=bf]{caption}
\DeclareCaptionLabelSeparator{pipe}{ $\mathbf{|}$ }
\usepackage{hyperref}
\usepackage[backend=bibtex,style=nature,sorting=none,maxnames=5]{biblatex}
\AtEveryBibitem{\clearfield{month}}
\AtEveryBibitem{\clearfield{day}}
\AtEveryBibitem{\clearfield{issn}}
\AtEveryBibitem{\clearfield{doi}}
\AtEveryBibitem{\clearfield{url}}
\AtEveryBibitem{\clearfield{arxivId}}
\AtEveryBibitem{\clearfield{eprint}}
\AtEveryBibitem{\clearfield{language}}

\usepackage[intlimits]{amsmath}
\usepackage{braket}
\usepackage{textcomp} 
\usepackage{authblk} 
\newcommand{\xzpf}{x_\mathrm{zpf}}
\newcommand{\meff}{m_\mathrm{eff}}
\newcommand{\Pin}{P_\mathrm{in}}
\newcommand{\kappaex}{\kappa_\mathrm{ex}}
\newcommand{\kappain}{\kappa_\mathrm{in}}
\newcommand{\kappaout}{\kappa_\mathrm{out}}
\newcommand{\oc}{\omega_\mathrm{c}}
\newcommand{\Om}{\Omega_\mathrm{m}}
\newcommand{\varone}[1]{\langle #1 \rangle}

\DeclareMathOperator{\Lor}{\mathcal{L}}

\newcommand{\sinamp}{s_{\mathrm{in}}}
\newcommand{\sout}{s_{\mathrm{out}}}
\newcommand{\Pout}{P_\mathrm{out}}

\newcommand{\arrtwo}[2]{\left[ \begin{array}{c} #1\\#2 \end{array} \right]}

\begin{document}
\title{Strong optomechanical interactions in a sliced photonic crystal nanobeam}
\author{Rick Leijssen}
\author{Ewold Verhagen\thanks{verhagen@amolf.nl} }
\affil{\small Center for Nanophotonics, FOM Institute AMOLF, Science Park 104, 1098 XG, Amsterdam, The Netherlands}
\date{}
  \maketitle
\begin{abstract}
\bfseries
Cavity optomechanical systems can be used for sensitive detection of mechanical motion and to control mechanical resonators, down to the quantum level.
The strength with which optical and mechanical degrees of freedom interact is defined by the photon-phonon coupling rate $\mathbf{g_0}$,
which is especially large in nanoscale systems.
Here, we demonstrate an optomechanical system based on a sliced photonic crystal nanobeam,
that combines subwavelength optical confinement with a low-mass mechanical mode.
Analyzing the transduction of motion and effects of radiation pressure we find a coupling rate $\mathbf{g_0/ 2 \pi\approx11.5}$~MHz, exceeding previously reported values by an order of magnitude.
Using this interaction we detect the resonator's motion with a noise imprecision below that at the standard quantum limit,
even though the system has optical and mechanical quality factors smaller than $\mathbf{10^3}$.
The broad bandwidth is useful for application in miniature sensors,
and for measurement-based control of the resonator's motional state.
\end{abstract}

\begin{refsection}
The motion of a mechanical resonator can be read out with extreme sensitivity in a suitably engineered system whose optical response is affected by the displacement of the resonator.
The resultant coupling between optical and mechanical degrees of freedom also gives rise to a radiation pressure force that enables actuation, tuning, damping, and amplification of the resonator, with applications ranging from classical information processing to quantum control of macroscopic objects\supercite{Aspelmeyer2014,Vanner2015}.
Such control can be established either passively, by employing the intrinsic dynamics of the system\supercite{Teufel2011,Chan2011,Verhagen2012}, or actively, by using the outcome of displacement measurements\supercite{Cohadon1999}.
Fast, sensitive measurement of nanomechanical displacement can as such be used for optical cooling\supercite{Cohadon1999,Wilson2014}, squeezed light generation\supercite{SafaviNaeini2013}, quantum non-demolition measurements\supercite{Purdy2013, Vanner2013} and enhancing sensor bandwidth\supercite{Gavartin2012,Poggio2007}.

In a cavity optomechanical system, which has an optical resonance frequency $\oc$ that depends on the position of a resonator, both the sensitivity of a displacement measurement and the magnitude of effects caused by radiation pressure forces are governed by two parameters: on the one hand the strength with which acoustic and optical degrees of freedom interact, expressed as the magnitude of the resonator's influence on the frequency $\oc$, and on the other hand the cavity linewidth $\kappa$.
The interaction strength is characterized at the most fundamental level by the vacuum optomechanical coupling rate $g_0$, as it enters the optomechanical interaction Hamiltonian $\hat{H}_\mathrm{int}=\hbar g_0 \hat{a}^\dag \hat{a}(\hat{b}+\hat{b}^\dag)$, where $\hat{a}$ and $\hat{b}$ are the photon and phonon annihilation operators, respectively.
As this Hamiltonian shows, $g_0$ is the frequency response of the optical cavity due to the mechanical displacement in a typical quantum state, where the total number of phonons is of the order of $1$.

Per photon in the cavity, the effective optomechanical measurement rate\supercite{Wilson2014,SafaviNaeini2013}, as well as the radiation-pressure induced alteration of a resonator's frequency and damping through dynamical backaction, scale with $g_0^2/\kappa$.
Improving this ratio is thus desirable for more sensitive measurements and for better optical control of the mechanical resonator.
Decreasing the optical damping $\kappa$ to a low value has been very fruitful, but can present several drawbacks as well: narrow linewidths place stringent demands on excitation sources and fabrication tolerances, and make integration of many devices, e.g. in practical sensor arrays, difficult.
Moreover, dynamical instabilities and nonlinear linewidth broadening limit the number of photons with which a high-Q cavity can be populated.
Finally, several schemes for measurement and control in fact rely on fast, broadband optical response\supercite{Vanner2011,Vanner2015,Genes2008}.
The photon-phonon coupling rate $g_0$, vice versa, is given by $g_0=G\xzpf$, where $G=\partial \oc / \partial x$ is the frequency shift per unit displacement $x$ and $\xzpf = \sqrt{\hbar/2 \meff \Om}$ are the zero-point fluctuations of a resonator with mass $\meff$ and frequency $\Om$.
The magnitude of $g_0$ is maximized in suitably engineered miniature systems, as $G$ and $\xzpf$ benefit from small cavity size and small resonator mass, respectively.
Indeed, the highest values of $g_0$ to date have been achieved in micrometer-size devices such as photonic crystal cavities\supercite{Eichenfield2009,Gomis-Bresco2014,Gavartin2011,Chan2011,Deotare2012,SafaviNaeini2013} or disk resonators\supercite{Ding2011,Balram2014}, with reported values ranging up to about $g_0/2\pi\approx 1$~MHz\supercite{Chan2012a,Balram2014}.

In this work, we show that optomechanical coupling rates can be significantly enhanced by using photonic modes with subwavelength confinement.
We realize a sliced photonic crystal nanobeam in which light is highly confined in a nanoscale volume near the moving dielectric interfaces of a low-mass resonator, leading to unprecedented interaction strengths.
We use a simple free-space optical setup to address the structure and demonstrate optical tuning of the mechanical resonance frequency, as well as sensitive readout of mechanical motion.
The observed optical forces and measurement sensitivity provide us with two independent ways to determine the vacuum coupling rate to be $g_0/2\pi\approx11.5$~MHz.
We demonstrate displacement readout with a detection imprecision below that at the standard quantum limit, i.e. with a noise level that is comparable to the quantum fluctuations of the resonator.
We achieve this using only $22$~\textmu W of detected power even in a system with modest optical and mechanical quality factors.
The operation with a relatively large cavity bandwidth is especially attractive for system integration and miniature sensor technologies as well as measurement-based control in nano-optomechanical systems.

\begin{figure*}
  \centering
  \includegraphics[width=0.8\textwidth]{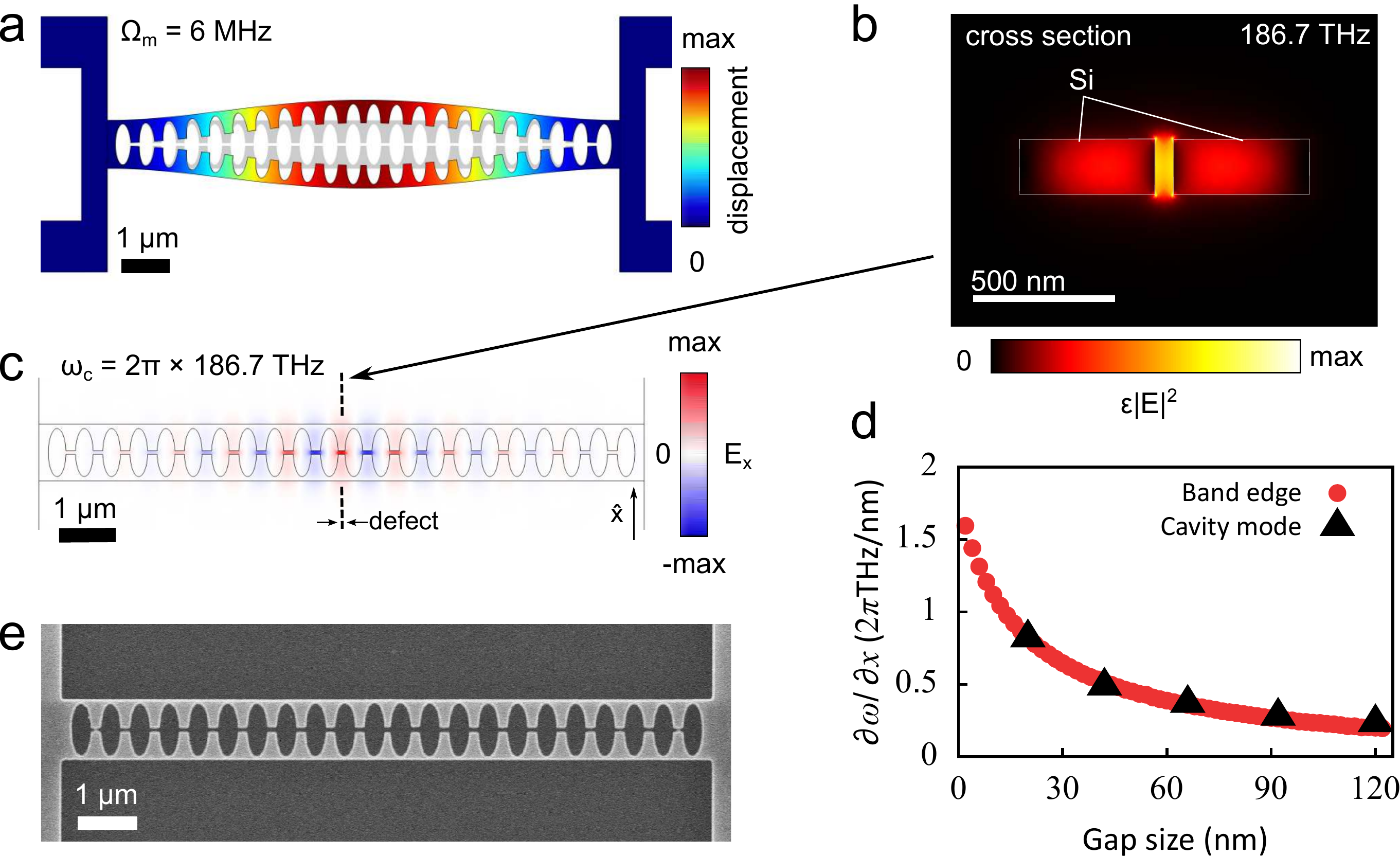}
  \caption{{\bfseries Geometry and resonances of the sliced nanobeam.} (\textbf{a}) Simulated displacement profile of the fundamental (in-plane) mechanical resonance of the structure.
  (\textbf{b}) Cross section in the center of the sliced nanobeam (indicated by the dashed line in \textbf{c}), showing the simulated energy density distribution of the fundamental optical cavity mode of the structure.
  (\textbf{c}) Simulated transverse electric field profile of the fundamental optical cavity mode of the structure.
  (\textbf{d}) Simulated frequency shift as a result of an outward displacement of $1$~nm. The cavity mode shift was determined by simulating the full nanobeam and introducing a uniform displacement along the beam.
  (\textbf{e}) Electron micrograph of a fabricated device.}
  \label{simandsem}
\end{figure*}

{\bfseries Working principle.} To realize a large photon-phonon coupling rate $g_0 = G \xzpf$, we develop a novel system that is based on a patterned silicon photonic crystal nanobeam, which combines optical confinement with flexural mechanical motion (Fig.~\ref{simandsem}).
The beam is `sliced' through the middle such that it mechanically resembles a pair of doubly clamped beams, coupled through the clamping points at the ends of the nanobeam.
Figure~\ref{simandsem}a shows the simulated fundamental in-plane mechanical resonance of the sliced nanobeam structure.
The small width ($80$~nm) of the narrowest parts of the half-beams ensures both the mass ($\approx 2.4$~pg) and the spring constant of the nanobeam are small, leading to large zero-point fluctuations $\xzpf$.

The in-plane mechanical motion causes a strong change in the separation distance $d$.
This leads to a giant change of the optical response, as we design the beam to concentrate light in the subwavelength slit separating the two halves.
In general, a displacement-induced frequency shift of an optical mode depends on the fraction of the energy density that is located near the moving dielectric boundaries\supercite{Johnson2002}.
To maximize this effect, we rely on the high localization of energy that can occur in systems with dielectric discontinuities with subwavelength dimensions\supercite{Robinson2005}, in this case provided by the narrow slit through the middle of the sliced nanobeam.
The periodic patterning of the beam creates a photonic crystal, with a quasi-bandgap for TE-polarized modes guided by the beam (see Supplementary Information).
The waveguide mode at the lower edge of the band gap has strongly confined electric fields in the nanoscale gap separating the `teeth' of the two half-beams (Fig.~\ref{simandsem}b).

The truly subwavelength character of this waveguide is revealed by calculating its effective mode area, which we suitably define as $A=\int \mathrm{d}V\,W(\mathbf{r}) / (a W_\mathrm{max})$, where the energy density $W(\mathbf{r})=\epsilon(\mathbf{r})|\mathbf{E}(\mathbf{r})|^2$ has its maximum $W_\mathrm{max}$ just at the vacuum side of the gap boundary, and we integrate over a full unit cell with period $a$.
The mode area is only $2.38\times 10^{-14}$~m$^2$ for a gap width of $60$~nm, or in other words $A=0.011 \lambda^2$, with $\lambda$ the wavelength in vacuum.
In fact, it is even 8 times smaller than the squared wavelength \emph{in silicon}, even though the maximum energy density is actually localized in the vacuum gap (Fig.~\ref{simandsem}b).
This subwavelength mode area is essential to the sliced nanobeam and makes it stand out with respect to other designs, including the related double-beam `zipper' cavity\supercite{Eichenfield2009, Gong2011, Deotare2012, SafaviNaeini2013}, where the optical cavity modes of two photonic-crystal nanobeams are coupled by placing the beams close together.

Recently it was shown that with a similar approach photonic crystal nanobeam cavities can be created that have a high quality factor and an ultrasmall mode volume\supercite{Ryckman2012,Seidler2013}.
We introduce a defect in the periodicity in the middle of the beam so confined cavity modes are created with a frequency in the bandgap.
These are derived from the band of interest (i.e. the lower bandgap edge) by reducing the width of the central pair of teeth, such that the effective refractive index is locally reduced.
Figure~\ref{simandsem}c shows the simulated field profile of the lowest-order optical cavity mode.

Numerical simulations confirm that the frequency of both the band edge and the defect cavity mode derived from it respond strongly to a displacement of the two half-beams, reaching $G=\partial\oc/\partial x \approx 2\pi\times 0.4$~THz/nm for a gap width of $60$~nm (Fig.~\ref{simandsem}d).
As expected, this value increases for smaller gap sizes, due to an increase of the fraction of the energy in the gap\supercite{Johnson2002,Robinson2005}.
We define the displacement coordinate as $x=d/2$, such that it can be directly related to the maximum lab-frame displacement of the antisymmetric mechanical mode depicted in Fig.~\ref{simandsem}a.
Note that the choice of the definition of $x$ is in principle arbitrary (with a properly matched definition of $\meff$), whereas the coupling rate $g_0$ is independent of this definition.
To determine the optical frequency shift, the entire half-beams are displaced in the simulation.
The displacement of the actual mechanical mode is not uniform along the beam (Fig.~\ref{simandsem}a), meaning that due to the finite extent of the optical mode the value of $G$ will be slightly reduced.
Taking into account the optical and mechanical mode profiles (Figs.~\ref{simandsem}a,c), we estimate it to be $0.90$ times the value shown in Fig.~\ref{simandsem}d (see Supplementary Information).

Using standard lithography techniques (see Methods for details), we realize sliced nanobeams in silicon with a length of $11$~\textmu m separated by an average gap size of $60$~nm.
An electron micrograph of a fabricated device is shown in Fig.~\ref{simandsem}e.

\begin{figure*}[t]
  \centering
  \includegraphics[width=0.8\textwidth]{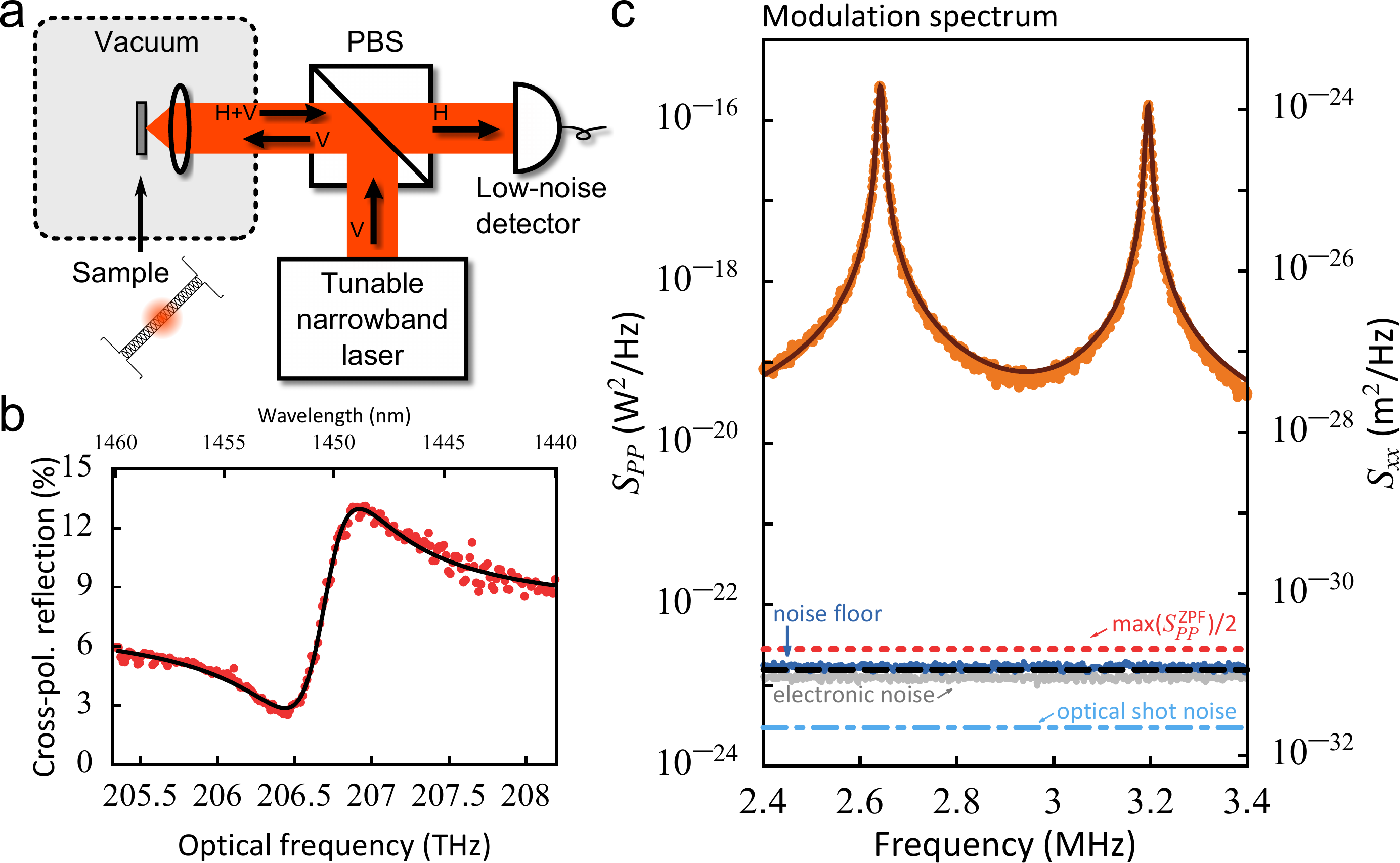}
  \caption{{\bfseries Free-space characterisation.} (\textbf{a}) Schematic diagram of the free-space readout method (PBS: polarizing beamsplitter; H,V: horizontally and vertically polarized light.
  See Methods for details).
  (\textbf{b}) Reflection spectrum (red datapoints) and fit with a Fano lineshape (black line).
  (\textbf{c}) Modulation spectrum of the reflected light obtained with the laser frequency on-resonance with the cavity (orange datapoints), and a fit of the two mechanical resonances (brown line).
  The noise floor (blue datapoints) was obtained by reflecting the laser light from the unpatterned substrate and matching the intensity on the detector.
  The black dashed line is the sum of the measured electronic noise (grey datapoints) and the optical shot noise calculated from the intensity on the detector (light blue dash-dotted line).
  The red dotted line shows the peak value of $S_{PP}^{\textrm{ZPF}} / 2$ for the lowest-frequency resonance, which we obtained from the fit of the measured thermal spectrum via the relation $S_{PP}^{\textrm{zpf}}(\Om) / 2 = S_{PP}^\mathrm{th}(\Om)\hbar\Om/4 k_B T$. }
  \label{setupspectra}
\end{figure*}

{\bfseries Free-space readout.} We address our structure using a simple reflection measurement, schematically shown in Fig.~\ref{setupspectra}a.
The employed resonant scattering technique\supercite{McCutcheon2005} places the sample between crossed polarizers to allow the detection of light scattered by the cavity mode (whose dominant polarization is oriented at 45\textdegree~to the polarizers) while suppressing light reflected by the substrate.
By scanning the frequency of a narrowband laser we record the reflection spectrum, depicted for one of the samples in Fig.~\ref{setupspectra}b.
The dispersive lineshape is caused by interference of the resonant scattering of the cavity with non-resonant scattering by the nanobeam.
The cross-polarized reflectance $R$ is thus well fitted by a Fano lineshape\supercite{Fan2003,Galli2009}:
\begin{equation}  \label{fanoeq}
  R (\Delta) = \left| c e^{i \varphi} - \frac{\sqrt{\kappain \kappaout }}{- i \Delta + \kappa / 2} \right|^2 ,
\end{equation}
where $c$ and $\varphi$ are the amplitude and phase of the non-resonant scattering, respectively, and $\Delta \equiv \omega - \oc$ is the detuning of the laser frequency $\omega$ from the cavity resonance with linewidth $\kappa$.
The rate at which light can couple to the cavity mode from the free-space input beam is given by $\kappain$, whereas $\kappaout$ is the rate at which the cavity decays to the radiation channels that are detected through the output analyzer.
In principle, these coupling rates can be unequal because the light emitted by the cavity has a spatial mode profile that differs from the Gaussian input beam.

Fitting equation~(\ref{fanoeq}) to the reflection spectrum yields the center frequency and linewidth of the cavity, as well as a value for $\kappaex \equiv \sqrt{\kappain \kappaout}$.
We determine $\kappaex \approx 0.29 \kappa$ and the optical quality factor $Q_{\textrm{opt}} = \oc / \kappa \approx 400$.
The measured $Q_\mathrm{opt}$ is 2--3 times lower than the simulated one, a discrepancy that we attribute to fabrication imperfections.

Thermal motion of the nanobeam $\delta x$ modulates the cavity frequency by $\delta \oc = (\partial \oc / \partial x) \delta x$.
This produces a change in detected power proportional to the derivative of the reflection spectrum: $\delta P = \Pin (\partial R / \partial \oc ) \delta \oc$.
Here we assumed the intracavity amplitude is instantaneously affected by the mechanical motion, which is justified since $\Om \ll \kappa$ (see Supplementary Information for the more general case).
Thus, the power spectral densities of $x$ and $P$ are related as
\begin{equation}\label{SppSxxeq}
  S_{PP}(\Omega) = \Pin^2 \left( \frac{\partial R}{\partial \oc} \right)^2 \left( \frac{\partial \oc}{\partial x} \right)^2 S_{xx}(\Omega) = \Pin^2 \left( \frac{\partial R}{\partial \Delta} \right)^2 \frac{g_0^2}{\xzpf^2} S_{xx}(\Omega).
\end{equation}

Figure~\ref{setupspectra}c shows the detected spectral density $S_{PP}$ for the laser tuned to the optical resonance frequency, with a relatively high optical power incident on the sample ($\Pin = 367$~\textmu W), corresponding to a detected power of $22$~\textmu W.
Because of the linear relation between $S_{PP}$ and $S_{xx}$ shown in equation~(\ref{SppSxxeq}), this is a direct measurement of the spectrum of thermal motion in the nanobeam.

The two peaks at $2.6$ and $3.2$~MHz correspond to the two fundamental in-plane modes of the coupled halves of the nanobeam.
For a perfectly symmetric structure, the coupling leads to two eigenmodes: a common mode, for which the half-beams move in phase and their separation $d$ is not affected, and a differential mode, for which anti-phase movement of the half-beams results in maximal variation of $d$.
Fabrication-related imperfections can break the symmetry of the system, such that the actual normal modes $\vec{\psi}_{\alpha,\beta}$ are linear combinations of the half-beam eigenmodes $\vec{\psi}_{1,2}$\supercite{Sun2012}: $\vec{\psi}_{\alpha} = A_\alpha ( \vec{\psi}_1 \sin \theta - \vec{\psi}_2 \cos \theta )$ and $\vec{\psi}_{\beta} = A_\beta (\vec{\psi}_1 \cos \theta + \vec{\psi}_2 \sin \theta )$, where $\theta$ can in principle take any value.
As we show in the Supplementary Information, the splitting between the mode frequencies $\Om^{\alpha,\beta}$ is enhanced due to the presence of compressive stress in the studied sample, which also reduces the mode frequencies with respect to the simulated value in absence of stress of $6$~MHz.
Since the two modes generally affect the separation $d$ differently, they have different photon-phonon coupling rates $g_0$, which are maximal for a purely differential mode ($\theta=\pi/4$).
With our definition $x=d/2$, this is reflected in the fact that the ratio between the zero-point fluctuation amplitudes of the normal modes is $\xzpf^\alpha / \xzpf^\beta = \sqrt{\Omega_\beta (1 + \sin 2 \theta) / \Omega_\alpha (1 - \sin 2 \theta)}$ (see Supplementary Information).
The variance in $x$ due to thermal motion in the two modes is set by the equipartition theorem, taking into account this difference in $\xzpf$.
The ratio between the areas of the two resonance peaks in the experimental spectrum of $S_{PP}$ therefore directly yields the mixing angle $\theta$.

In fact, fitting two resonant modes to the displacement spectrum also allows determining the transduction factor that relates the measured optical power spectral density $S_{PP}$ to the displacement spectrum $S_{xx}$.
To do so, we calculate the thermal variance $\varone{x^2}_\mathrm{th}=2 \xzpf^2 k_B T/\hbar\Om$.
We determine $\xzpf$ from the measured $\theta$ and from the effective mass of purely antisymmetric motion, which we computed from the simulated displacement profile to be $\meff \approx 0.39 m$, with $m$ the total mass of the beam.
We further assume that the temperature $T$ of the mechanical bath is equal to the lab temperature.
The validity of this assumption is tested by performing power- and detuning-dependent measurements presented in the Supplementary Information.
The resulting scale for the displacement spectral density $S_{xx}$ is shown on the right side of Fig.~\ref{setupspectra}c.
Note that the chosen convention of $x$ allows directly comparing the readout of the two mechanical resonances on this scale.

To determine the sensitivity with which the displacement spectrum of the beam can be read out, we consider the detection noise floor for the measurement shown in Fig.~\ref{setupspectra}c, which is composed of electronic noise of the photodetector and the optical shot noise of the detected light.
Their measured combined imprecision (blue datapoints in Fig.~\ref{setupspectra}c) is over 7 orders of magnitude smaller than the measured signal.

A general assessment of the sensitivity capabilities of the measurement is made by comparing the detection noise imprecision to the (shot noise) imprecision $S_{xx}^{\textrm{imp}} (\Om)$ of a resonator read out at the standard quantum limit (SQL)\supercite{Anetsberger2009,Teufel2009}.
The imprecision at the SQL is equal to half of the spectral density of the zero-point fluctuations $S_{xx}^\mathrm{zpf} (\Om) / 2$.
We determine this value from the measured thermal noise spectrum of the lowest-frequency mode via the average phonon occupancy of the mechanical mode $k_B T / \hbar \Om$, and indicate it in Fig.~\ref{setupspectra}c with the red dotted line.
The optical shot noise of the light impinging on the detector, and even the total measurement noise floor, are lower than the imprecision noise at the SQL.

Readout of a nanomechanical resonator with an imprecision below that at the SQL was first achieved in 2009\supercite{Teufel2009,Anetsberger2009} making use of high-quality optical and mechanical modes.
These high quality factors were instrumental because the ability to perform a measurement with SQL-level sensitivity scales, per intracavity photon, with the single-photon cooperativity $C_0 = 4 g_0^2 / \kappa \Gamma$.
This shows it depends on the photon-phonon coupling strength as well as the optical linewidth $\kappa = \oc / Q_\mathrm{opt}$ and the mechanical linewidth $\Gamma = \Om / Q_\mathrm{m}$.
The fact that here we achieve a detection noise imprecision below that at the SQL with optical and mechanical quality factors of both less than 500 attests to the large optomechanical coupling strength, and could have important application in broadband, sensitive nanoscale sensors.

{\bfseries Determining the photon-phonon coupling rate.}
To quantify the optomechanical interaction strength in the fabricated devices, we model the transduction of thermal displacement fluctuations using equation~(\ref{SppSxxeq}) and use it to fit a low-power measurement on a structure for various laser detunings.
We do this by calculating the variance of the optical power fluctuations $\delta P$ at the detector resulting from displacement fluctuations $\delta x$ of a mechanical mode with known (thermal) variance.
Integrating equation~(\ref{SppSxxeq}) over a single mechanical mode and using our expression for the reflection spectrum $R(\Delta)$ (equation~(\ref{fanoeq})), yields
\begin{equation}\label{varp2eq}
  \langle P^2 \rangle = 8 \Pin^2 g_0^2 \frac{k_B T}{\hbar \Om} \frac{  \kappaex^2 \left( \Delta \kappaex - c \Delta \kappa \cos (\varphi) - c (\Delta^2 - \kappa^2 / 4) \sin (\varphi) \right)^2 }
       {(\Delta^2 + \kappa^2 / 4)^4  } ,
\end{equation}
which is independent of the choice of the displacement coordinate $x$.

\begin{figure}
  \centering
  \includegraphics[width=0.45\textwidth]{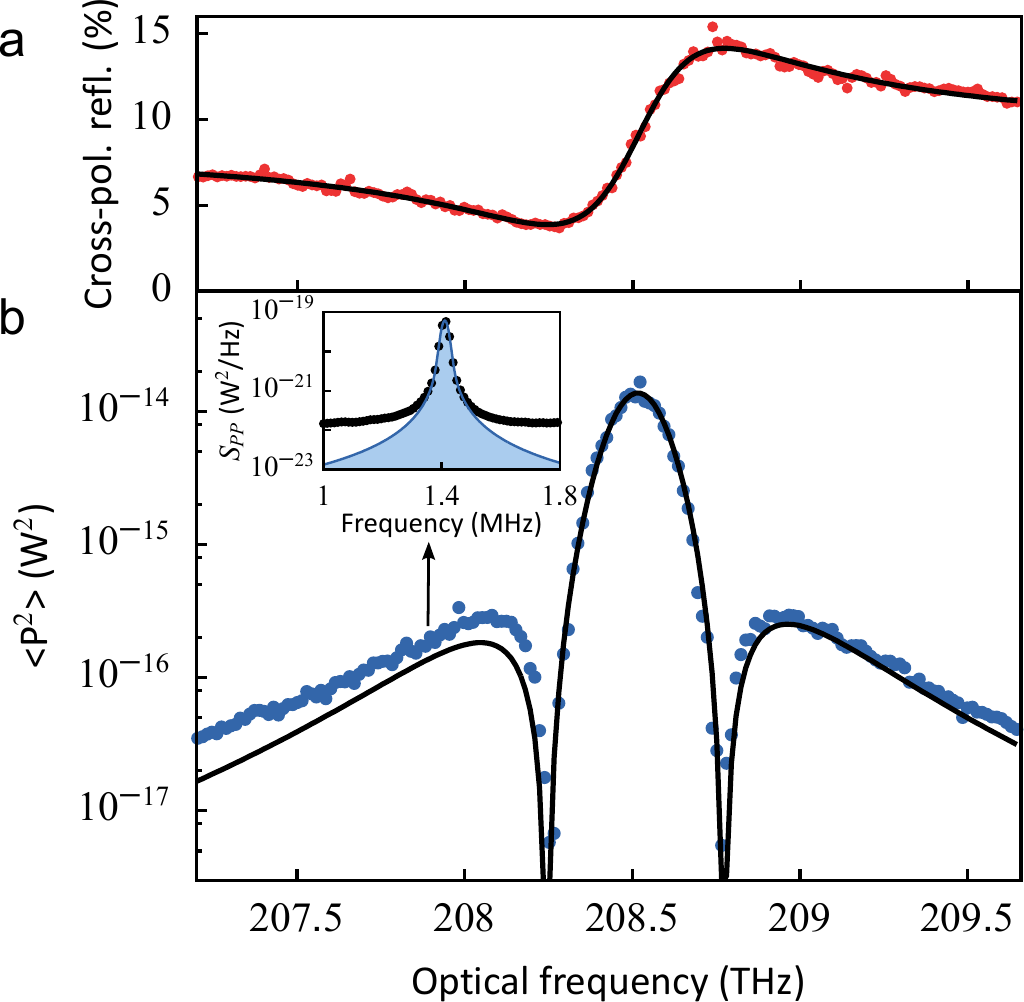}
  \caption{{\bfseries Sensitivity of the displacement measurement.} (\textbf{a}) Reflection spectrum (red datapoints) and fit with Fano lineshape (black line) for this particular nanobeam. (\textbf{b}) Detected optical variance in the fundamental mechanical resonance (blue datapoints) with power incident on the sample $\Pin = 8.5$~\textmu W.
  The datapoints were obtained by fitting the fundamental mechanical resonance peak in the measured modulation spectra (inset).
  The signal originates from thermal motion so it varies only with the sensitivity of the measurement.
  The black line shows our model, which uses the parameters obtained from the reflection spectrum in \textbf{a} and is fitted to the data to determine $g_0 = 11.5$~MHz.}
  \label{transd}
\end{figure}

The measured variance of the optically modulated signal due to the lowest-frequency mechanical mode is shown in Fig.~\ref{transd}b.
The variance is minimal when the derivative of the reflection signal (Fig.~\ref{transd}a) vanishes.
Interestingly, due to the dispersive lineshape the transduction is largest for the laser tuned to resonance.
The line shown in Fig.~\ref{transd}b is a fit of equation~(\ref{varp2eq}) to the data, using only $g_0$ as a free fitting parameter (all other parameters having been determined in independent measurements). The corresponding value for $g_0 / 2 \pi$ is $11.5$~MHz, which is an order of magnitude larger than previously reported values\supercite{Chan2012a,Balram2014,Chan2011,SafaviNaeini2013,Eichenfield2009,Gomis-Bresco2014,Ding2011}.

To compare this photon-phonon coupling rate to the prediction from our simulation we estimate the zero-point fluctuations of the structure.
Using the measured mechanical resonance frequency and the simulated effective mass, we obtain $\xzpf = \sqrt{\hbar/2 \meff \Om} \approx 0.08$~pm for a purely anti-symmetric mode.
With the simulated frequency response $G$, this yields a prediction of $g_0 / 2 \pi \approx 26$ MHz.
To take into account the observed asymmetry of the mechanical mode, we should apply a correction factor of $0.76$, based on our knowledge of $\theta$ (see Supplementary Information).
This results in an expected value of $g_0 / 2 \pi \approx 20$~MHz.
We attribute the remaining discrepancy to fabrication imperfections, that could result in a different overlap of the optical and mechanical modes than simulated.
So in fact, these simulations show that a further increase of $g_0$ even beyond the measured value is possible.

\begin{figure}
  \centering
  \includegraphics[width=0.45\textwidth]{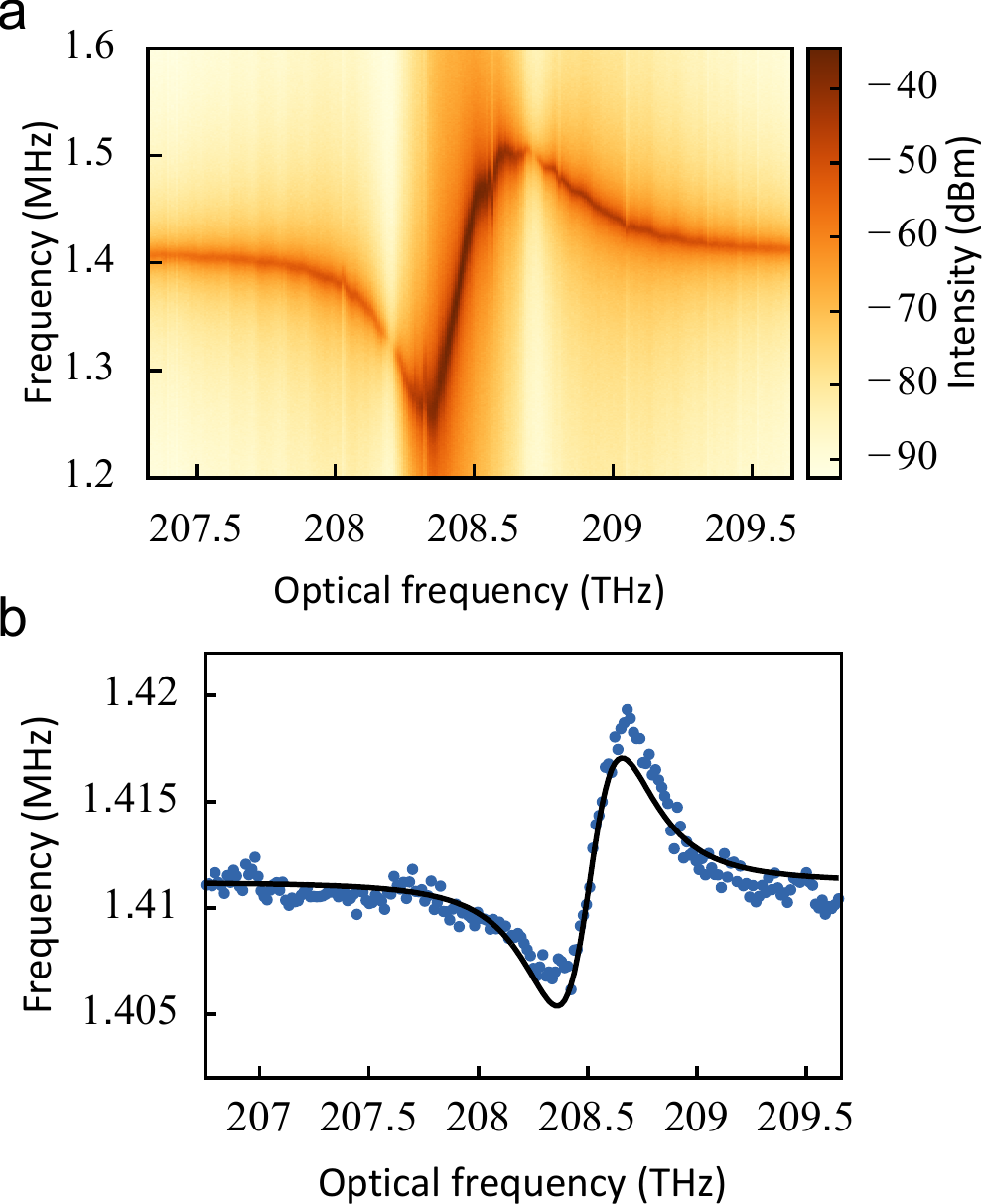}
  \caption{{\bfseries Optical tuning of the mechanical resonance frequency.} (\textbf{a}) Spectrogram showing optical tuning of mechanical resonance frequency with power incident on the sample $\Pin = 140 $~\textmu W. (\textbf{b}) Fitted frequency of the fundamental mechanical resonance (blue datapoints) with $\Pin = 8.5$~\textmu W. The black line shows a fit using the model for optical spring tuning (equation~(\ref{springeq})).}
  \label{spring}
\end{figure}

{\bfseries Optical spring tuning.} While we tune the laser frequency across the optical resonance a pronounced shift of the mechanical resonance frequency is observed. In Fig.~\ref{spring}a this is shown for the same structure we studied in Fig.~\ref{transd}.
This well-known optical spring effect is caused by the radiation pressure force being opposed to (aligned with) the mechanical restoring force when the laser is detuned below (above) the resonance frequency, changing the effective spring constant and therefore the mechanical resonance frequency\supercite{Aspelmeyer2014}.
The equation that describes this behaviour in the limit of a large cavity linewidth ($\kappa \gg \Om$) is
\begin{equation}  \label{springeq}
    \delta \Om = g_0^2 \frac{2 \Delta}{\Delta^2 + \kappa^2 / 4}  \frac{\kappain \Pin}{\hbar \omega}.
\end{equation}

From equation~(\ref{springeq}) we recognize that the optical spring tuning shown in Fig.~\ref{spring} provides a second, independent way to characterize the photon-phonon coupling rate.
Figure~\ref{spring}b shows the center frequency of the mechanical resonance extracted from the same measurement as the variances in Fig.~\ref{transd}b, as well as a fit using equation~(\ref{springeq}).
To estimate $g_0$ from this fit we need to know $\kappain$, which we cannot easily determine as it generally depends on the overlap between the focused Gaussian beam and the cavity mode profile.
However, we can find bounds for $\kappain$ by considering the total decay rate $\kappa$ and $\kappaex = \sqrt{\kappain\kappaout}$, which were determined from the fit to the reflection spectrum.
On the one hand we know $\kappain \leq \kappaex$, i.e. the collection efficiency is at least as efficient as the overlap with a Gaussian beam, and on the other hand $\kappain \geq 2 \kappaex^2 / \kappa$, i.e. at most half of the light escaping from the cavity can be collected because of the vertical symmetry of the structure.
Combining these bounds with the fit of the optical spring effect yields a range for $g_0$ between 10 and 13 MHz, in good agreement with the value obtained from the analysis of measurement transduction.
The fact that the spring shift can be fully explained by the radiation pressure force as predicted by equation~(\ref{springeq}) shows that forces due to photothermoelastic effects\supercite{Metzger2008} are likely insignificant compared to radiation pressure.

{\bfseries Nonlinear transduction.}
As a consequence of the large photon-phonon coupling rate, the thermal motion of the nanobeam induces frequency changes that are appreciable with respect to the linewidth of the cavity, which results in nonlinear transduction.
This generates spurious signals at integer multiples of, and combinations of, the strongest modulation frequencies.
Detection of such signals at multiples of the mechanical resonance frequency resulting from thermal motion was reported previously\supercite{Lin2009,Deotare2012,Cohen2013} and compared to quadratic optomechanical coupling\supercite{Doolin2014}.

\begin{figure}[t]
  \centering
  \includegraphics[width=0.45\textwidth]{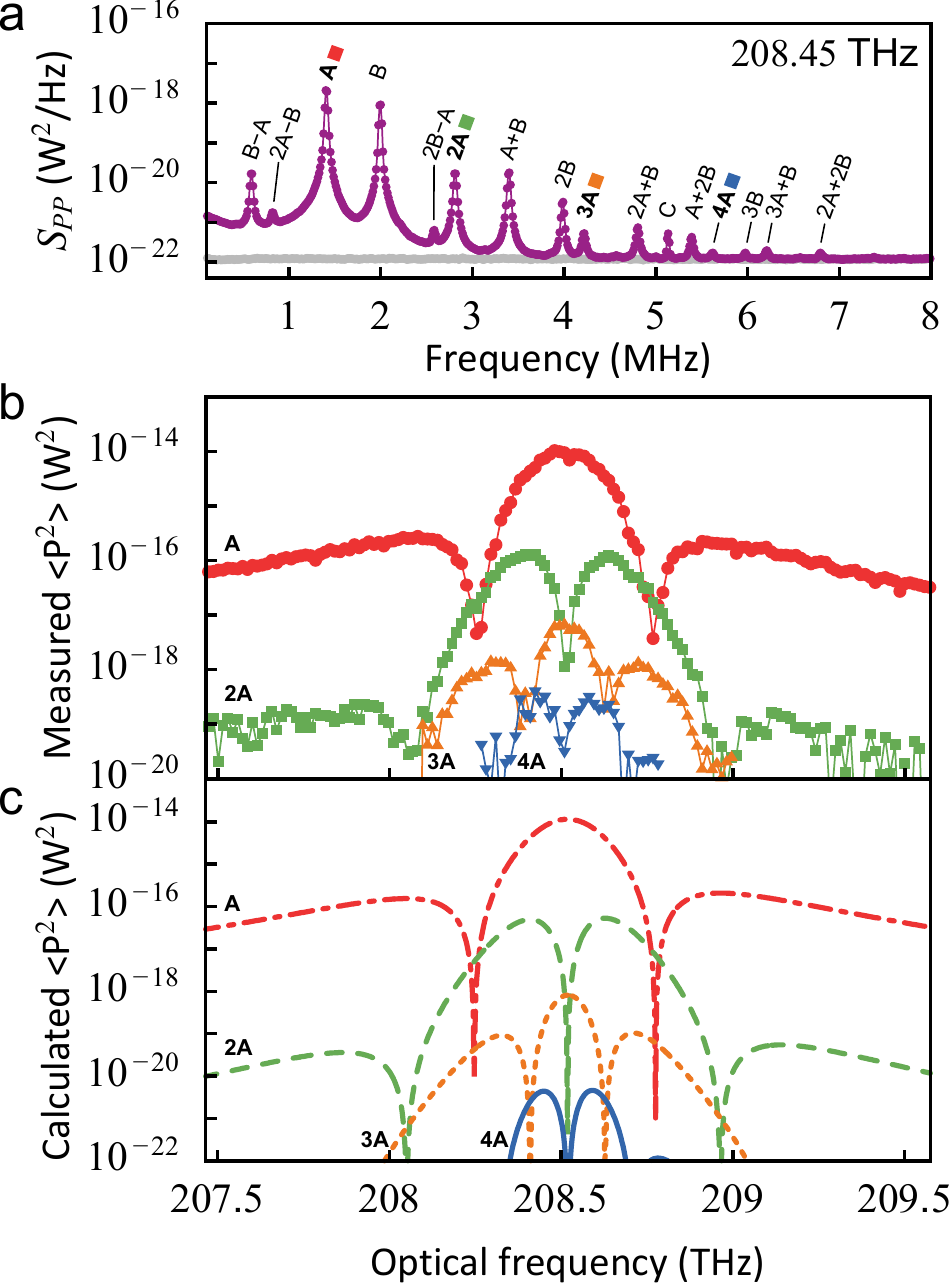}
  \caption{{\bfseries Nonlinear transduction.} (\textbf{a}) Measured mechanical spectrum of the sliced nanobeam (purple datapoints), using near-resonant light with power incident on the sample $\Pin = 8.5$~\textmu W.
  Three peaks A, B and C are mechanical resonances of the nanobeam, all other visible peaks correspond to integer multiples and combinations of frequencies A and B.
  The electronic noise floor is shown with grey datapoints.
  (\textbf{b}) Areas under the peaks corresponding to integer multiples of frequency A, obtained by fitting the peaks in the spectrum.
  (\textbf{c}) Calculated variance of the reflected signal, using the experimentally obtained parameters.}
  \label{nonlineartransd}
\end{figure}

Figure~\ref{nonlineartransd}a shows a transduced spectrum where we identify 15 peaks as integer multiples and combinations of the two fundamental mechanical resonances at $1.4$~MHz (``A'') and $2.0$~MHz (``B''): $\Omega_{j,k} = | j \textrm{A} \pm k \textrm{B} |$, with $j,k \in \{0,1,2,\dots \}$.
Peaks corresponding to different order $(j + k)$ have a different detuning dependence, but all peaks with the same order differ only by a constant factor.
To illustrate the detuning dependence of the higher-order peaks, we
plot the variance of the peaks $j \textrm{A}$ for $j = \{1,2,3,4\}$ in Fig.~\ref{nonlineartransd}b.
The detected height of the higher-order peaks can be predicted by a Taylor expansion of the amount of light in the cavity around the average detuning\supercite{Doolin2014} (see Supplementary Information), the result of which is shown in Fig.~\ref{nonlineartransd}c.

Note that the higher-order peaks in this calculation were not fitted to the data, but follow from the value of $g_0$ we obtained by fitting the first-order peak, as shown in Fig.~\ref{transd}.
The measured nonlinear sidebands are larger than expected (corresponding to a suggested increase of $g_0$ of about $60 \%$). The origin of this discrepancy is unknown. Possible explanations include higher-order optomechanical coupling\supercite{Doolin2014} or mechanical nonlinearities\supercite{Ramos2014}.
However, the symmetry and shape of the curves match the experimental data, which confirms that the detuning dependence corresponds to the successive derivatives of the reflection spectrum (Fig.~\ref{transd}a).

{\bfseries Discussion.} The free-space readout method we employ provides an easy and robust way of coupling light to the cavity.
We have intentionally engineered the cavity defect such that it has a significant dipole moment\supercite{Quan2010},
allowing coupling to free space at an appreciable rate.
This makes it unnecessary to create an explicit loss channel for coupling, e.g. in the form of a grating or feeding waveguide.
Moreover, the Fano-shape of the reflection spectrum allows direct transduction of motion to optical amplitude modulation for a laser tuned to the cavity resonance (where dynamical radiation pressure backaction is zero), without more complicated interferometric schemes.
As a result of the efficient coupling to free space, the bandwidth of the cavity is large (0.5 THz), which is appealing in the context of applications that require frequency matching of multiple systems: together with the small system footprint, it could assist the integration of such optomechanical transducers in sensor arrays\supercite{Thijssen2014} or effective optomechanical metamaterials\supercite{Heinrich2011a}.

Of course, for applications that benefit from enhanced measurement sensitivity such as measurement-based control of the mechanical quantum state, it could be worthwhile to realize a higher optical quality factor by introducing tapering along the nanobeam\supercite{Quan2010,Ryckman2012}.
To simultaneously allow efficient free-space coupling would in such a case require special attention, in the form of tailoring the spatial mode profile of the cavity radiation.
This could be especially important for effects that depend on the intracavity photon number, such as the demonstrated optical spring effect, as the rates $\kappain$ and $\kappaout$ will differ.
Further quantification of their individual magnitudes (e.g. through systematic variation of incident and detected mode profiles) will thus be valuable.

Likewise, we expect that the mechanical quality factor for the nanobeams we employ can be improved with suitable design principles and optimization of the fabrication process.
Indeed, measurements on similar-sized silicon nanobeams and cantilevers suggest that quality factors in the range of $10^4$ to $10^5$ should be possible at room and cryogenic temperature, respectively\supercite{SafaviNaeini2013,Tao2014}.
Nonetheless, we point out that because of our large coupling rate, even with the current modest values of both optical and mechanical quality factors the single-photon cooperativity in this structure reaches $C_0 = 0.16$. The value of this quantity, which compares optomechanical coupling strength and dissipation, and is for example a measure for the capability of the system to perform measurements at the SQL, is on par with many recently reported systems with much
higher quality factors\supercite{Aspelmeyer2014}.

In conclusion, we demonstrated an optomechanical device with a large photon-phonon coupling rate $g_0 / 2 \pi = 11.5$~MHz, and used it to demonstrate sensitive measurement of nanomechanical motion and pronounced optical tuning of the mechanical resonance frequency.
It is interesting to note that the regime of large coupling rate and modest optical linewidth is beneficial in the context of achieving strong mechanical tuning, as parametric instability is suppressed.
We revealed that the working mechanism relies on an optical mode with a subwavelength mode area.
We predict this approach can be extended to yield even larger coupling rates, or to be applied to modes with higher mechanical frequencies.
In the current device the photon-phonon coupling rate $g_0$ exceeds the mechanical resonance frequency $\Om$, which is one of the requirements for ultrastrong coupling\supercite{Aspelmeyer2014, Nunnenkamp2011, Yeo2014}.
With further improvements in both the coupling rate and the optical quality factor, the present approach might provide a route to simultaneously reach $g_0 > \Om$ and $g_0 \approx \kappa$.
It will be interesting to explore to what extent this regime can be used to exploit nonlinear optomechanical interactions at the single-photon level.

{\bfseries Acknowledgements.} The authors thank R. Thijssen for valuable discussions. This work is part of the research programme of the Foundation for Fundamental Research on Matter (FOM), which is part of the Netherlands Organisation for Scientific Research (NWO).
E.V. gratefully acknowledges an NWO-Vidi grant for financial support.

\section*{Methods}

{\bfseries Numerical simulation.} All numerical eigenmode simulations were performed using finite-element software COMSOL Multiphysics.
In mechanical simulations, the connection between the substrate and the support pads was modeled as a fixed boundary, while all other boundaries were kept free. To find the guided modes of the photonic crystal nanobeam, a unit cell was simulated with Floquet boundary conditions along the propagation direction and in the other directions perfect electric conductors at several micrometers distance from the structure. Finally, to simulate the cavity mode, a full nanobeam including support pad was modeled with perfectly matched layers on all sides, again at several micrometers distance from the beam.

{\bfseries Fabrication.} The structures were fabricated from a silicon-on-insulator wafer with a device layer thickness of $200$~nm, and a buried oxide layer of $1$~\textmu m thick.
A resist layer of hydrogen silsesquioxane (HSQ) with a thickness of $80$~nm was spincoated on top and patterned using electrons accelerated with $30$~kV.
The resist was developed using TMAH and then the pattern was transferred to the silicon layer using a reactive-ion etch process with SF$_6$/O$_2$ gases, optimized for anisotropy and selectivity.
To release the structures, the oxide layer was dissolved in a $20 \%$ HF solution.
After this step the structure was dried with a critical point dryer to prevent the sliced beams being pulled together during the drying process.
The suspension of the nanobeams from their support pads was designed to allow some relief of compressive stress along the beam.
The compressive stress is present in most free-standing structures created from SOI\supercite{Yamashita2015}, but it has a large effect for our structures because of their low stiffness.

{\bfseries Free-space setup.} The laser beam (New Focus Velocity 6725) was focused on the sample by an aspheric lens with a numerical aperture of $0.6$.
A polarizing beamsplitter provided a cross-polarized detection scheme, where any light that was directly reflected was rejected and only light that coupled to the sample, placed at $45$\textdegree, was transmitted to the detector.
Both the lens and the sample were in a vacuum chamber to reduce mechanical damping by air molecules.
All experimental results shown were performed with a pressure of about $4$~mbar, except the spectra in Fig.~\ref{setupspectra}, where the pressure was lower than $10^{-3}$~mbar.
At a pressure of $4$~mbar, the mechanical quality factor was lowered to approximately $200$.

{\bfseries Analysis of modulated reflection signals.}
We detected the reflection signal using a low-noise InGaAs-based photoreceiver (Femto HCA-S) and analyzed it using an electronic spectrum analyzer (Agilent MXA).
We fitted the peaks in the modulation spectra using a Lorentzian convolved with a Gaussian distribution, also called a Voigt lineshape.
The Gaussian contribution accounted for the resolution bandwidth of the spectrum analyser, as well as for frequency-noise broadening at relatively large optical input power $\Pin$.
With high $\Pin$, small fluctuations in incoupling efficiency or laser intensity thermally shifted the cavity resonance, which resulted in frequency noise via optical spring tuning of the mechanical resonance.

\printbibliography
\end{refsection}

\section*{Supplementary Information}

\setcounter{figure}{0}
\renewcommand{\thefigure}{S\arabic{figure}}
\setcounter{equation}{0}
\renewcommand{\theequation}{S\arabic{equation}}

\begin{refsection}

\section*{Model for transduction}
We derive the effect of a small frequency modulation of the intracavity field at the output of a cavity, which allows us to set up a model that describes the transduction of thermal motion to optical intensity modulation.
This derivation follows the calculation shown by \textcite{Gorodetsky2010}, with the important difference that we include the non-resonant contribution in the reflection spectrum that leads to the Fano lineshape we observe.

We start from the equations describing the behaviour of the optomechanical system
\begin{equation}
  \begin{aligned}
    \dot{a} = (i \left(\Delta - G x(t)\right) - \kappa / 2) a(t) + \sqrt{\kappain} \sinamp (t), \\
    \ddot{x} (t) + \Gamma_\mathrm{m} \dot{x}(t) + \Om^2 x (t) = - \hbar G | \bar{a}|^2,
  \end{aligned}
\end{equation}
where $a$ is the internal field in the cavity, $\sinamp$ is the input field related to the input power $\Pin = \hbar \omega | \sinamp |^2$, $\Delta$ is the detuning of the input light from the cavity resonance $\oc$, $\kappa$ is the cavity decay rate, $\kappain$ is the coupling rate to the input channel and $G = \partial \oc / \partial x$ is the optomechanical frequency response.
The frequency $\Om$ and damping rate $\Gamma_m$ of the mechanical resonator are influenced by the number of photons in the cavity $| \bar{a}|^2$, an effect we neglect in the following by assuming a low input power. This simplification is motivated by the fact that we seek to predict the amplitude of the mechanically-induced light modulation, not the frequency of such modulations. Moreover, dynamical backaction affecting mechanical linewidth is small in devices that have large $\kappa/\Om$.

We consider a small harmonic oscillation of the mechanical resonator $x(t) = x_0 \cos(\Om t)$, which causes a modulation of the cavity frequency with amplitude $x_0 G$, or a modulation of the optical intracavity phase with amplitude $x_0 G / \Om$. If the modulation is small ($x_0 G \ll \kappa$), this yields
\begin{equation}
  \begin{aligned}
    a_x &= \sinamp \sqrt{\kappain} \Lor (0) \\
    & \hspace{1em} \times \left( 1 - \frac{i}{2} x_0 G \Lor (\Om) e^{-i \Om t} - \frac{i}{2} x_0 G \Lor (- \Om) e^{i \Om t} \right), \\
    \Lor (\Omega) &= \frac{1}{-i (\Delta + \Omega) + \kappa / 2}.
  \end{aligned}
\end{equation}
The cavity is coupled to the output $\sout$ with the coupling rate $\kappaout$. The equation for the output field reads
\begin{equation}\label{inputoutput}
    \sout = c e^{i \varphi} \sinamp - \sqrt{\kappaout} a_x,
\end{equation}
where the first term is caused by the nonresonant scattering from the input to the output with amplitude $c$ and phase $\varphi$.

In the experiment, we measure the intensity $| s_\mathrm{out}|^2$ and feed it to a spectrum analyser, which yields the single-sided spectrum of the signal.
Since we are interested in the strength of the spectral component at the mechanical oscillation frequency $\Om$, we highlight the time dependence here:
\begin{equation}
    |\sout|^2 (t) = c^2 |\sinamp|^2
    + \kappaout |a_x (t)|^2
    - c \sqrt{\kappaout}
    (e^{i \varphi} \sinamp a_x^* (t) + e^{-i \varphi} \sinamp^* a_x (t) ).
\end{equation}
The first term is constant so it does not contribute to a signal at $\Om$. Substituting $a_x$ into the other two terms, and discarding any terms not oscillating at $\pm \Om$ yields
\begin{equation}
  \begin{aligned}
    |s_\mathrm{out}|^2 (t) \bigg\rvert_{\pm\Om}
   &= \frac{1}{2} x_0 G |s_{in}|^2 \sqrt{\kappa_{in} \kappa_{out}} \\
    & \hspace{1em} \times \bigg( \sqrt{\kappa_{in} \kappa_{out}} |\Lor(0)|^2 \left( \left[ i e^{i \Om t} \left( \Lor^* (\Om) - \Lor (-\Om) \right) \right] + \mathrm{c.c.} \right) \\
    & \hspace{2em} - c \left( \left[ i e^{i \Om t} \left( e^{i \varphi} \Lor^*(0) \Lor^* (\Om) - e^{-i \varphi} \Lor(0) \Lor(- \Om ) \right) \right] + \mathrm{c.c.} \right) \bigg)
  \end{aligned}.
\end{equation}
This expression contains the modulation amplitude.
In our experiment, we compare the variance of the modulation to the known variance of the mechanical thermal motion $\varone{x^2}_\mathrm{th} = 2 \xzpf^2 k_B T / \hbar \Om$.
Therefore we calculate the variance of $\Pout = \hbar \omega_0 | s_{out} |^2$ due to the modulation at $+ \Om$ and $- \Om$, which will both contribute to the signal at $+ \Om$ in the single-sided spectrum.
We can write $\Pout \Big\rvert_{\pm \Om} = A e^{i \Om t} + A^* e^{-i \Om t}$, which leads to $\langle |\Pout |^2 \rangle_{\Om} = 2 |A|^2$. After some algebra, we arrive at
\begin{equation}
  \begin{aligned}
    \langle |\Pout |^2 \rangle_{\Om}
    &= \frac{2 x_0^2 G^2 P_{in}^2 (\kappa_{in} \kappa_{out})}
       {(\Delta^2 + \kappa^2 / 4)^2 ( ( \Delta + \Om)^2 + \kappa^2/4) ( (\Delta - \Om)^2 + \kappa^2 / 4 )} \\
    & \hspace{1.5em} \times \Big[ \Delta^2 \kappa_{in} \kappa_{out}
      - 2 \Delta c \sqrt{\kappa_{in} \kappa_{out}} ( \Delta \kappa \cos \varphi + (\Delta^2 - \kappa^2 / 4 ) \sin \varphi) \\
    & \hspace{3em} + c^2 \bigg( \Delta^2 ( \Om^2 + \kappa^2 ) \cos^2 (\varphi) \\
    & \hspace{6.5em} +  (\Delta^4 - \Delta^2 \kappa^2 / 2 + \kappa^2 \Om^2 / 4 + \kappa^4 / 16) \sin^2 (\varphi) \\
  & \hspace{8em} - 2 \cos (\varphi) \sin (\varphi) (- \Delta^3 \kappa + \Delta \kappa^3 / 4 + \Delta \kappa \Om^2 / 2 )  \bigg) \Big].
  \end{aligned}
\end{equation}

If we evaluate this expression in the bad-cavity limit ($\Om \ll \kappa $), we find it is directly related to the derivative of the Fano lineshape $\partial R / \partial \Delta$:
\begin{equation}
  \begin{aligned}
    \langle |\Pout|^2 \rangle_{\Om}
  &= \frac{2  x_0^2 G^2 \Pin^2 \kappain \kappaout \left( \Delta \sqrt{\kappain \kappaout} - c \Delta \kappa \cos (\varphi) - c (\Delta^2 - \kappa^2 / 4) \sin (\varphi) \right)^2 }
       {(\Delta^2 + \kappa^2 / 4)^4  } \\
       &= \frac{1}{2} x_0^2 G^2 \Pin^2 \left(\frac{\partial R}{\partial \Delta} \right)^2.
  \end{aligned}
\end{equation}

We note that imperfect transmission of the optics between the sample and the detector scales the detected signal in the same way as the input power $\Pin$, and will enter the equations in the same way.
Finally, we substitute the variance due to the modulation amplitude $x_0$ by the variance of the mechanical motion: $x_0^2 G^2 \rightarrow 2 \varone{x^2}_\mathrm{th} G^2 = 4 g_0^2 k_B T / \hbar \Om$, which leads to equation~\ref{varp2eq}~in the main text.

\section*{Nonlinear transduction}
The previous section started from the assumption that the frequency modulation $\delta \oc = G \delta x $ is small with respect to the cavity linewidth, $\delta \oc \ll \kappa$, and considered only the resulting linear transduction at the modulation frequency $\Om$.
In this section we show that the first signature of large $\delta \oc$ is the appearance of nonlinear transduction, which produces a signal at multiples of the modulation frequency $\Om$.
For the second-order transduction, this was shown by \textcite{Doolin2014}, where also a quadratic optomechanical coupling was taken into account.
Here we derive the result for any higher-order terms of nonlinear transduction.

In the non-resolved sideband regime ($\kappa \gg \Om$), the optical fields in the cavity reach a steady state much faster than the timescale of mechanical motion.
The intracavity amplitude can then be written as
\begin{equation}
a (t) = \frac{\sqrt{\kappain} \sinamp}{-i (\Delta - \delta \oc (t) ) + \kappa / 2},
\end{equation}
which combined with equation~(\ref{inputoutput}) yields
\begin{equation}\label{Rfuncu}
    \frac{|\sout|^2}{|\sinamp|^2} = c^2
    + \frac{4 \kappain \kappaout}{\kappa^2} \frac{1}{1 + u^2}
    - \frac{2 c \sqrt{\kappain \kappaout}}{\kappa}
    \frac{e^{i \varphi} (1- iu) + e^{-i \varphi} (1 + iu)}{1+u^2}.
\end{equation}
Here we defined $u \equiv 2 ( \Delta - \delta \oc (t) ) / \kappa$, which implies u is detuning- and time-dependent.
We now summarize equation~(\ref{Rfuncu}) as $R'(u)$ and find the Taylor expansion for small $\delta \oc$ around $u_0 \equiv 2 \Delta / \kappa$:
\begin{equation}\label{RTaylor}
R'(u) = R'(u_0) - \frac{2 \delta \oc}{\kappa} \frac{\partial R'(u_0)}{\partial u}
+ \ldots + \frac{( - 2 \delta \oc / \kappa )^n}{n!} \frac{\partial^n R'(u_0)}{\partial u^n},
\end{equation}
where the last term depicts the $n$th order in the Taylor expansion.
We take a harmonic modulation of the cavity frequency $\delta \oc (t) = A \cos \Omega t$.
To leading order, $\delta \oc^n \approx A^n \cos (n \Omega t) / 2^{n-1}$.
This means that each successive term in the Taylor expansion in equation~(\ref{RTaylor}) gives the amplitude of a term at different frequency.

In the optomechanical system, the variance of the frequency modulations at the mechanical frequency $\Om$ is given by $\varone{\delta \oc^2} = G^2 \varone{x^2}_\mathrm{th} = 2 g_0^2 k_B T / \hbar \Om$.
Therefore we get the same variance in $R$ if we set $A = 2 g_0 \sqrt{k_B T / \hbar \Om}$.
We note that $\frac{\partial^n R'(u_0)}{\partial u^n} = (\kappa / 2)^n \frac{\partial^n R(\Delta)}{\partial \Delta^n}$, which for the variance of the signal at $n \Om$ leads to
\begin{equation}
\frac{\varone{P^2}}{\Pin^2} \bigg\rvert_{n \Om} = \varone{R^2} \Big\rvert_{n \Om}
= \frac{2 (g_0^2 k_B T / \hbar \Om)^n}{n!^2} \left( \frac{\partial^n R(\Delta)}{\partial \Delta^n} \right)^2.
\end{equation}
For $n=1$, the result of equation~\ref{varp2eq}~in the main text is again reproduced.
The result of a calculation of the variances of $n=1$--$4$, for the parameters used in the experiment, is shown in the main text in Fig.~\ref{nonlineartransd}c.

\section*{Waveguide modes in the sliced nanobeam}
\begin{figure}
\includegraphics[width=0.8\textwidth]{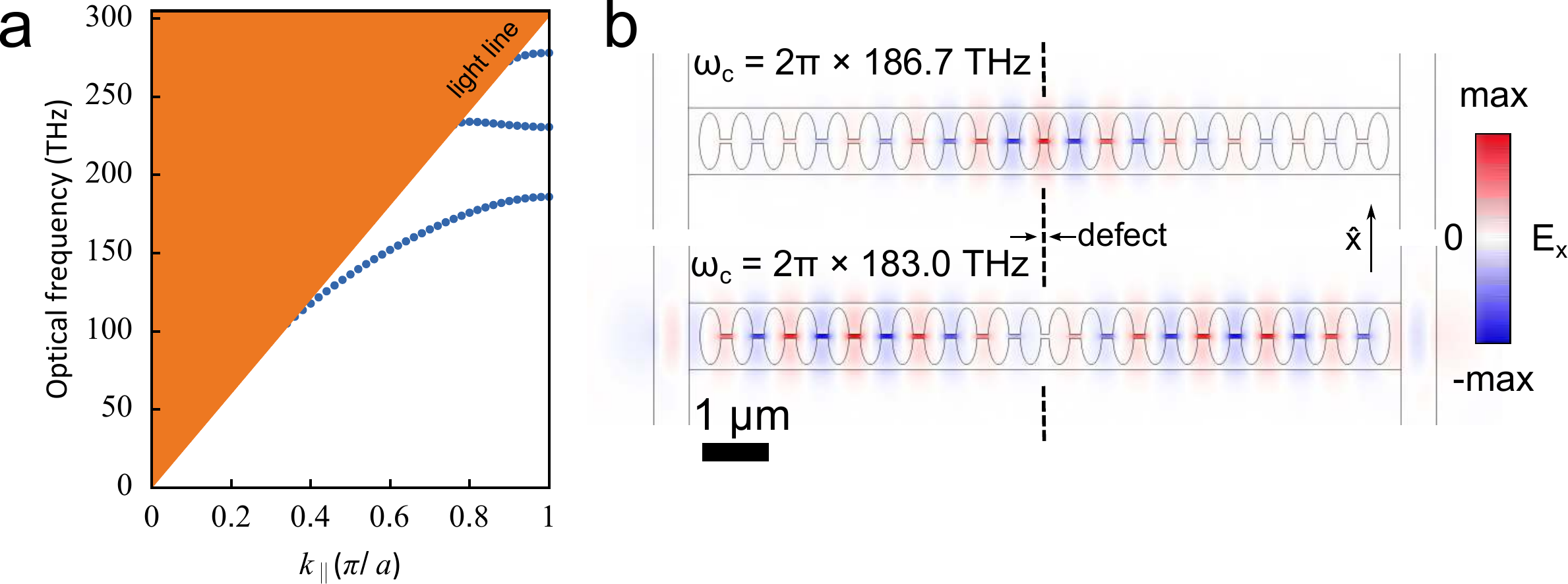}
\caption{{\bfseries Optical modes in the sliced nanobeam.} (\textbf{a}) Simulated dispersion diagram showing the TE-like waveguide modes in the periodic part of the sliced nanobeam.
A bandgap is opened by the periodic structure of elliptical holes around a frequency of 200 THz.
  (\textbf{b}) Simulated transverse electric field profiles of the first two cavity modes of the structure.
  The defect that is responsible for the creation of these modes is a slightly smaller distance between the two holes in the center of the beam.
  }
\label{waveguideandcavmodes}
\end{figure}

In this section we discuss the waveguide modes in the periodic region of the sliced nanobeam in more detail.
The free-standing silicon nanobeam acts as a waveguide, which can guide light via total internal reflection.
In this waveguide, the elliptical holes form a photonic crystal that opens a bandgap for modes with transverse electric (TE)-like symmetry (Supplementary Figure~\ref{waveguideandcavmodes}a).
This is not a full bandgap, since TM-like waveguide modes exist in the gap region.
If the symmetry of the structure is broken by fabrication imperfections, light in the bandgap region for TE-like modes can scatter to the TM-like modes and propagate along the nanobeam.
For this reason, this is sometimes referred to as a quasi-bandgap.

The guided mode at the lower edge of the band gap has the largest concentration of energy in the nanoscale gap in the middle of the beam.
The fact that a significant portion of this mode's energy is located in vacuum increases its frequency in comparison to a non-sliced nanobeam, which reduces the frequency width of the bandgap.
To ensure maximum mirror strength, the transverse size of the holes is made as large as possible.
The elliptical hole shape is as such important to realize a strong bandgap, in addition to providing favorable mechanical properties as mentioned in the main text.

To create optical cavity modes that are derived from the lower band edge, the defect is a local decrease in distance between two elliptical holes, which decreases the local effective refractive index and creates defect states in the bandgap region.
Supplementary Figure~\ref{waveguideandcavmodes}b shows the first two cavity modes created in this way in the sliced nanobeam.
Note that the higher-order cavity modes have a lower frequency, since they are less confined near the defect, so that the frequency is closer to that of the waveguide mode in the periodic structure.

\section*{Overlap of optical and mechanical mode profiles}

\begin{figure}
\centering
\includegraphics[width=0.45\textwidth]{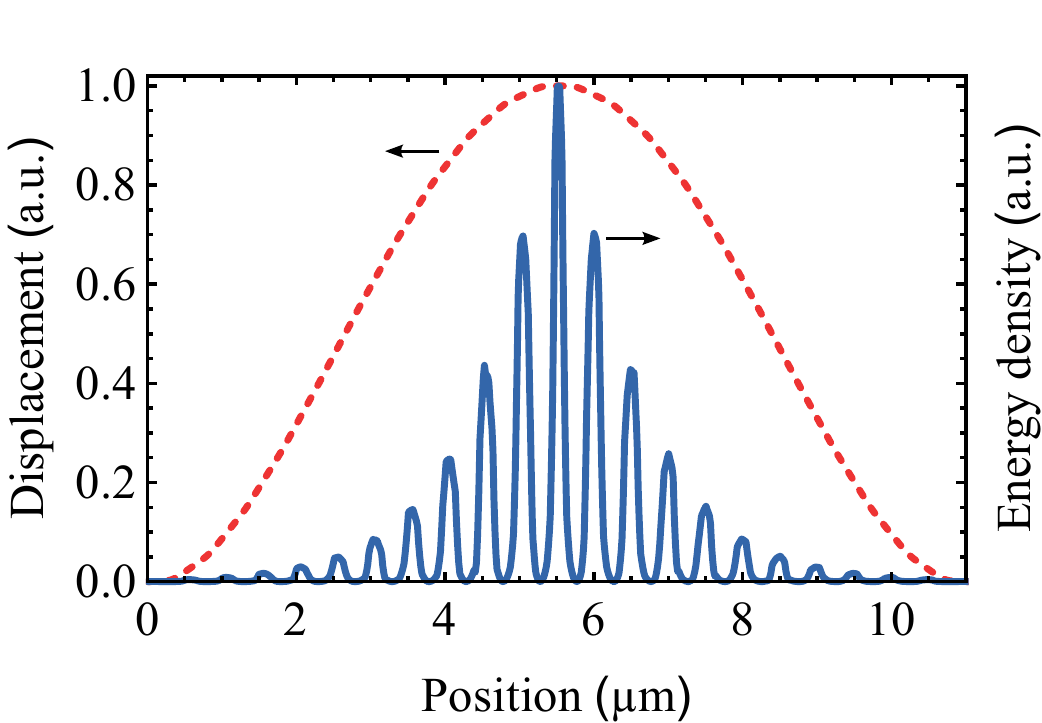}
\caption{Normalized resonant mode profiles of the fundamental optical and mechanical resonance of the sliced nanobeam. The red dotted line shows the displacement along the length of the beam, while the blue solid line represents the local energy density.}
\label{overlap}
\end{figure}

The frequency shift of the optical resonance of the sliced nanobeam due to mechanical motion depends on the overlap between the optical and mechanical mode profiles.
We simulated the frequency shift of the optical resonance of the sliced nanobeam due to a uniform mechanical shift of $1$~nm, as shown in the main text in Fig.~1\ref{simandsem}d.
To estimate the influence of the finite extent of the mode profiles on the response, we extract these profiles from the numerical simulation.
Supplementary Figure~\ref{overlap} shows the normalized displacement profile of the mechanical resonance and the normalized electromagnetic energy density profile of the optical cavity mode, as a function of the position along the nanobeam.

The mechanical mode profile closely resembles the mode profile of the fundamental mode of a doubly-clamped beam.
From the simulated displacement profile we calculate the effective mass for this purely antisymmetric motion to be $\meff \approx 0.39 m$, where $m$ is the total mass of the sliced nanobeam, which is indeed very close to the value obtained from the analytical displacement profile of a uniform doubly-clamped beam\cite{Cleland2003}.

The optical mode profile clearly shows the localization of the energy density in the small gaps between the silicon `teeth' of the structure.
The simple defect we introduce in the center of the beam localizes the optical cavity mode there, while the field decays exponentially away from the defect, where the optical frequency lies inside the photonic quasi-bandgap.

We compute the correction on the frequency shift due to the finite extent of the modes from the overlap integral between these two mode profiles, and find a factor of $0.90$ with respect to the frequency shift for a uniform displacement of the beam.

The higher-order cavity modes created by the defect are less strongly confined along the length of the beam, as shown in Supplementary Figure~\ref{waveguideandcavmodes}b.
Since the fundamental mechanical resonance has the largest displacement in the center of the beam, the higher-order cavity modes are less sensitive to this motion than the fundamental optical resonance.

\section*{Mechanical mode coupling}
The mechanical modes observed in a two-beam system are combinations of the motion of the individual beams, with in the perfectly symmetrical case fully in-phase or out-of-phase motion.
Here we derive the consequences of imperfect symmetry for the ratio of scattered power between the two modes\cite{Thijssen2015}.

Harmonic motion of the two beams at a certain frequency $\Omega$ can be described as:
\begin{equation}
\arrtwo{x_1(t)}{x_2(t)} =
\arrtwo{\psi_1}{\psi_2} \cos \Omega t \equiv \vec{\psi} \cos \Omega t \text{, where } \vec{\psi} = \arrtwo{\psi_1}{\psi_2}.
\end{equation}
Thus, $\psi_1$ and $\psi_2$ are the amplitudes of the oscillatory motion of the two individual beams, such that their variance is $\langle \psi_{1,2}^2 \rangle =\frac{1}{2}\psi_{1,2}^2$.

The optical response of the system is determined by the change in the distance between the beams: $d=x_1-x_2$.
Here we define $x$ as $x \equiv d/2$, which leads to the same value of $G = \partial \oc / \partial x$ for both mechanical modes.
Note that this choice of $x$ corresponds to the lab-frame displacement of the two beams if they move in antiphase.
The variance of $x$ due to harmonic motion described by $\vec{\psi}$ is then:
\begin{equation}
\varone{x^2}_\psi = \tfrac{1}{8}\left(\psi_1^2+\psi_2^2-2\psi_1\psi_2\right).
\end{equation}

The state vectors of the two normal modes can be written without loss of generality as
\begin{equation}
\begin{aligned}
\vec{\psi}_\alpha &= A_\alpha \arrtwo{\cos\theta}{\sin\theta}, &\vec{\psi}_\beta = A_\beta \arrtwo{\sin\theta}{-\cos{\theta}}.
\end{aligned}
\end{equation}
For both of these modes, we can calculate the variance of $x$, denoted as $\varone{x^2}_\alpha$ and $\varone{x^2}_\beta$, respectively:
\begin{equation}\label{varstheta}
\begin{aligned}
\varone{x^2}_\alpha
&=\frac{1}{8} A_\alpha^2(\cos^2\theta+\sin^2\theta+2\sin\theta\cos\theta)\\
&=\frac{A_\alpha^2}{8}(1+\sin{2\theta}),\\
\varone{x^2}_\beta &= \frac{A_\beta^2}{8}(1-\sin{2\theta})
\end{aligned}
\end{equation}

For the beams undergoing thermal motion, the variance is given by the equipartition theorem:
\begin{equation}\label{varsthermal}
\begin{aligned}
\varone{x^2}_\alpha &= \frac{k_B T}{m_\alpha \Omega_\alpha^2}, &\varone{x^2}_\beta &= \frac{k_B T}{m_\beta \Omega_\beta^2},
\end{aligned}
\end{equation}
where $m_\alpha$ and $m_\beta$ are the effective mass of these modes.
As shown in the previous section, for the differential mode the simulated effective mass with respect to the displacement coordinate $x$ is $\meff = 0.39 m$, with $m$ the total mass of the sliced nanobeam.
Evaluating equations~(\ref{varstheta}) and (\ref{varsthermal}) for a differential mode ($\theta = \pi / 4$) yields $A^2_\alpha = 4 k_B T / \meff \Omega^2_\alpha$ and similarly for $A^2_\beta$.
Substituting this back into equation~(\ref{varstheta}), we arrive at
\begin{equation}\label{varstotal}
\begin{aligned}
\varone{x^2}_\alpha &= \frac{k_B T ( 1 + \sin 2 \theta)}{\meff \Omega_\alpha^2}, &\varone{x^2}_\beta &= \frac{k_B T ( 1 - \sin 2 \theta)}{\meff \Omega_\beta^2}.
\end{aligned}
\end{equation}
We note that thermal variance is related to the zero-point fluctuations $\xzpf$ as
\begin{equation}
\varone{x^2}_\psi = 2 \frac{k_B T}{\hbar \Omega_\psi} (\xzpf^{\psi})^2, \text{ so } \xzpf^{\psi} = \sqrt{\frac{\hbar ( 1 \pm \sin 2 \theta)}{4 \meff \Omega_\psi}},
\end{equation}
where $\psi = \alpha,\beta$ and $+$ respectively $-$ is chosen as the sign for the term $\sin 2 \theta$.

Finally, we measure $\varone{P^2}$ and wish to relate this to a displacement variance $\varone{x^2}$.
We calculated $\meff$ using the simulated mode profile and assume the thermal bath temperature of the mechanical modes $T$ is equal to the lab temperature, which leaves only $\theta$ and the transduction factor $\partial P / \partial x$ unknown.
By measuring the area of both peaks $\varone{P^2}_\alpha$ and $\varone{P^2}_\beta$, we resolve the remaining ambiguity, allowing us to calibrate the displacement spectrum.

\section*{Influence of compressive stress and experimental disorder}
The simulation of the ideal structure shown in the main text predicts the resonance frequency of the fundamental in-plane resonance to be $6$~MHz, and additionally the frequency difference between the anti-symmetric and symmetric mode is negligible.
We expect the frequency of the out-of-plane resonances to be larger than that of the fundamental in-plane mode, as the narrowest part of the half-beams ($80$~nm) is smaller than the Si slab thickness ($200$~nm).

In our experiment we find significantly smaller values of $2.6$ and $3.2$~MHz  for one structure, and $1.4$ and $2.0$~MHz for a second structure. The differences between the frequencies of the two in-plane resonances are also larger than expected.
To find the origin of this effect we measure the experimentally observed disorder produced by the fabrication process of the second structure using a scanning electron microscope.
We then introduce these dimensions into the simulation, which results in slightly different eigenfrequencies.
However, the resulting values are still near $6$~MHz and additionally the two fundamental modes still have only a small frequency difference.

\begin{figure}
\centering
\includegraphics[width=0.5\textwidth]{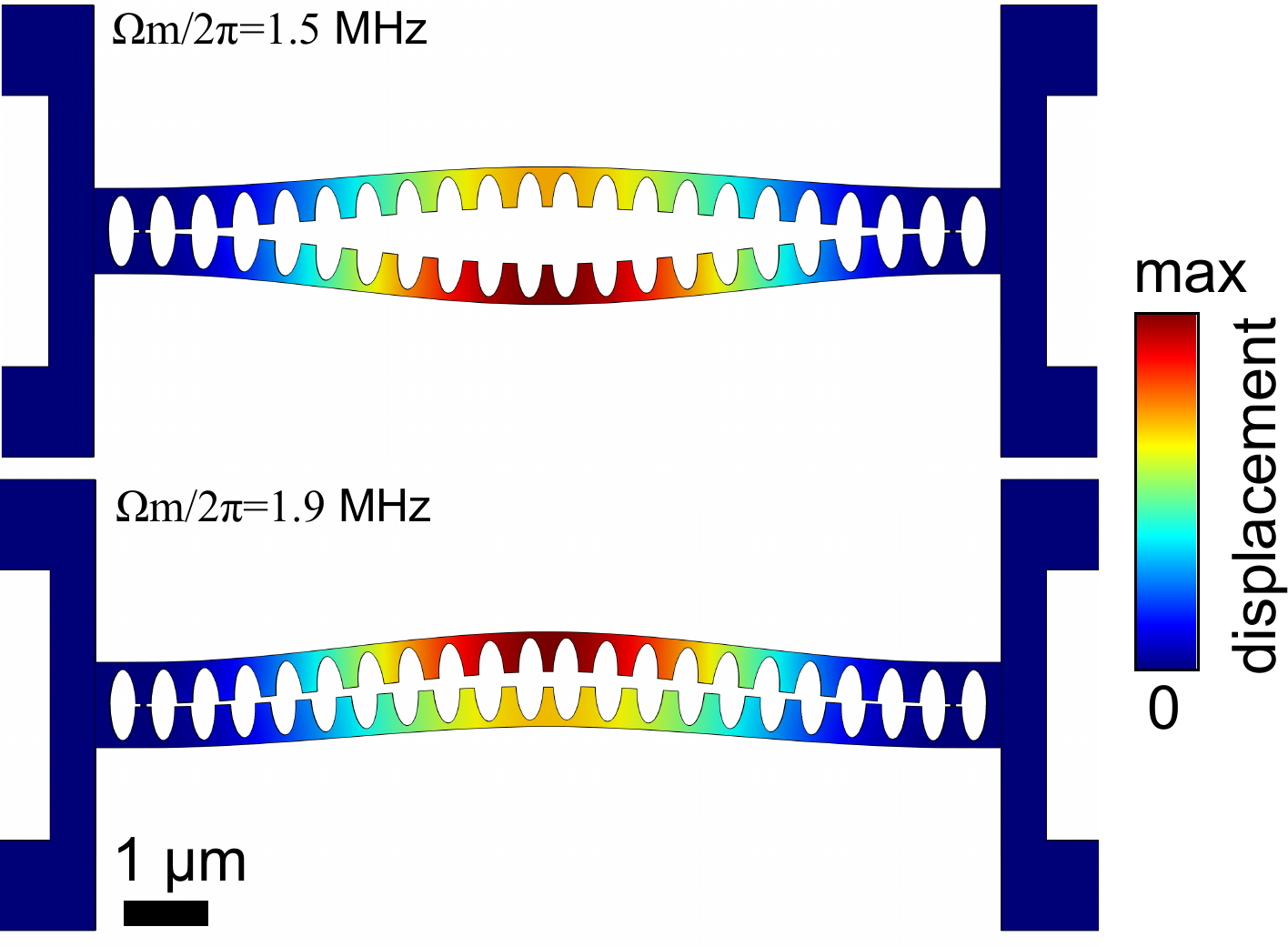}
  \caption{{\bfseries Simulation including disorder and compressive stress.} Simulated mechanical displacement profiles of the two fundamental in-plane resonances.
  The dimensions of the beam were matched to the realized dimensions using measurements with a scanning electron microscope, including differences in hole and gap size along the beam.
  In addition, a compressive stress was introduced in the simulation by displacing one of the support pads by $10$~nm along the direction of the beam.}
\label{stress}
\end{figure}

Finally, we add an extra step to the simulation to include a static compressive stress, which can result from the stress in the buried oxide layer of the silicon-on-insulator layer structure we use\cite{Yamashita2015}.
We define an initial displacement for one of the support pads along the length of the beam, which creates the stress.
Increasing the amount of displacement, and therefore the stress, decreases the eigenfrequencies predicted by the simulation.
This is in accordance with the theoretical expectation\cite{Bokaian1988}.
At the same time the difference between the eigenfrequencies of the two fundamental modes increases.

Supplementary Figure~\ref{stress} shows the displacement profiles that result from such a simulation.
At a displacement value of $10$~nm ($0.09 \%$ of the length of the beam), the resonances occur at frequencies very close to the experimentally measured values, at $1.5$ and $1.9$~MHz.

\section*{Influence of optical input power}
In the measurements shown in the main text, we use up to $370$~\textmu W of optical power incident on the nanobeam.
Using the parameters of our fit to the reflection spectrum, we estimate that this results in an intracavity intensity that corresponds to a maximum of $\approx 1000$ photons.
For this intensity in the cavity, we do not expect an increase in the cavity temperature of more than a few Kelvin.
As a first confirmation of this, we see thermal shifts of the cavity resonance frequency of less than $1$~nm, which corresponds to a temperature increase of less than $10 \%$.

\begin{figure}[tb]
\centering
\includegraphics[width=0.5\textwidth]{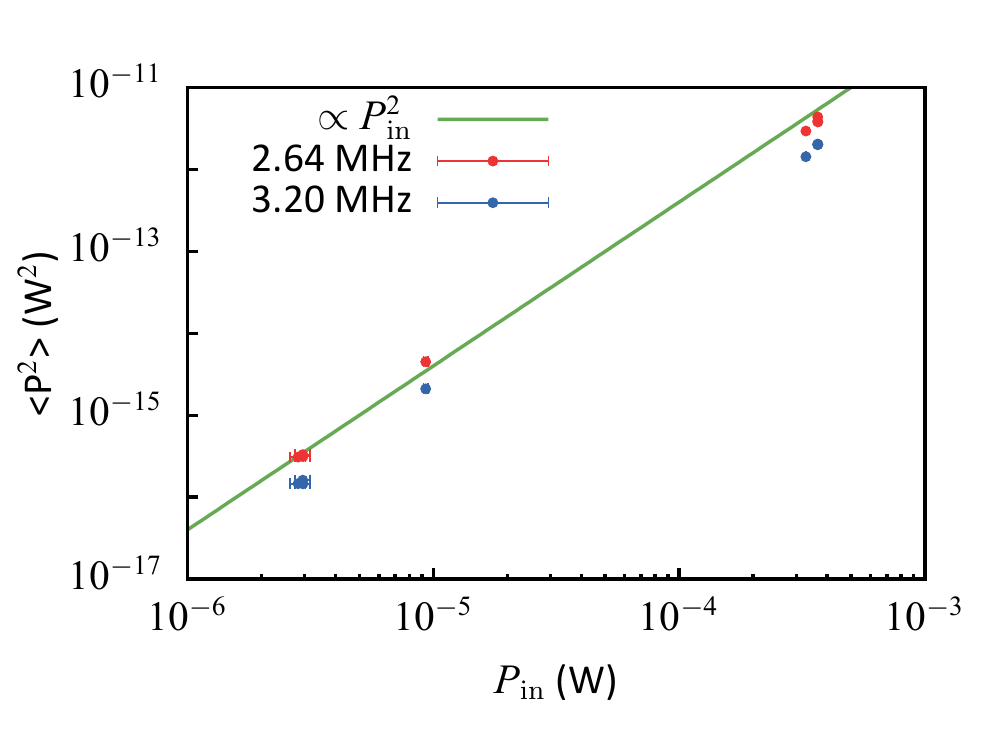}
\caption{{\bfseries Power dependence.} The red and blue datapoints show the measured signal from the two fundamental mechanical resonances, obtained by fitting both mechanical resonances in the spectra (see Fig.~2 in the main text for such a spectrum).
The data was taken at various input powers with resonant laser light (zero detuning).
The green line is a guide to the eye, with a slope corresponding to quadratic dependence on input power.
The errorbars indicate readout error of the input power but don't take into account possible variations in incoupling efficiency due to slight changes in alignment.}
\label{pwrdep}
\end{figure}

To check the assumption more thoroughly we measured the signal strength of the two fundamental mechanical frequencies as a function of power incident on the structure.
Supplementary Figure~\ref{pwrdep} shows the result for resonant light, both from the zero-detuning point in swept measurements and from individual measurements where the detuning was set to zero by minimizing the optically induced shift of the mechanical frequency.
Comparison with the line shows that the datapoints closely follow the expected quadratic dependence on input power for both peaks in the spectrum.
The largest source of uncertainty in this measurement are differences in coupling efficiency between the incoming laser beam and the cavity due to small changes in alignments, which can influence the signal strength between measurements.

\printbibliography
\end{refsection}

\end{document}